\documentclass[reprint,aps,prl,showpacs,superscriptaddress]{revtex4-1}
\usepackage{graphicx,epsfig}
\usepackage{epstopdf}
\usepackage{bm}

\newcommand{\exc}{He$_2^*$}
\begin{document}

\title{Excimers He$_2^*$ as Tracers of Quantum Turbulence in $^4$He in the {\itshape T}=0 Limit}
\author{D. E. Zmeev}
\affiliation{School of Physics and Astronomy, The University of Manchester, Manchester M13 9PL, UK}
\affiliation{Department of Physics, Lancaster University, Lancaster LA1 4YB, UK}
\author{F. Pakpour}
\affiliation{School of Physics and Astronomy, The University of Manchester, Manchester M13 9PL, UK}
\author{P. M. Walmsley}
\affiliation{School of Physics and Astronomy, The University of Manchester, Manchester M13 9PL, UK}
\author{A. I. Golov}
\affiliation{School of Physics and Astronomy, The University of Manchester, Manchester M13 9PL, UK}
\author{W. Guo}
\affiliation{Mechanical Engineering Department, Florida State University, Tallahassee, FL 32310-6046,
USA}
\author{D. N. McKinsey}
\affiliation{Department of Physics, Yale University, New Haven, CT 06520-8120, USA}
\author{G. G. Ihas}
\affiliation{Department of Physics, University of Florida, Gainesville, FL 32611-8440, USA }
\author{P. V. E. McClintock}
\affiliation{Department of Physics, Lancaster University, Lancaster LA1 4YB, UK}
\author{S. N. Fisher}
\affiliation{Department of Physics, Lancaster University, Lancaster LA1 4YB, UK}
\author{W. F. Vinen}
\affiliation{School of Physics and Astronomy, University of Birmingham, Birmingham B15 2TT, UK}

\begin{abstract}
We have studied the interaction of metastable $^4$He$_2^*$ excimer molecules with quantized vortices in superfluid $^4$He in the zero temperature limit. The vortices were generated by either rotation or ion injection. The trapping diameter of the molecules on quantized vortices was found to be $96\pm6$\,nm at a pressure of 0.1\,bar and $27\pm5$\,nm at 5.0 bar. We have also demonstrated that a moving tangle of vortices can carry the molecules through the superfluid helium. 
\end{abstract}

\maketitle

Turbulence, the complex dynamics of systems with many degrees of freedom on a broad range of lengthscales, is common in nature. Its understanding is important for both fundamental science and technology. A special case is the hydrodynamics of superfluid liquids in the limit of zero temperature \cite{Vinen06}, which, while behaving as an ideal fluid, has a quantum  constraint: vorticity is concentrated along the filamentary cores of quantized vortex lines, and the velocity circulation around any such line is equal to $\kappa=h/m_4=1.00\times10^{-3}$\,cm$^2$s$^{-1}$ (where $h$ is the Planck constant and $m_4$ is the mass of a $^4$He atom).  Turbulence in such a system, known as Quantum Turbulence (QT), is a dynamic tangle of vortex lines. 

The characterization of classical turbulence is a formidable task: all the velocity field has to be visualized at once, and the most important regions are those of enhanced vorticity. Usually small passive tracers of flow are used \cite{Nobach2009}. The case for visualization of QT is different. To begin with, velocity tracers cannot be used as they are not entrained by the superfluid. On the other hand, small particles are attracted to the cores of quantized vortices, in which they can be trapped and then traced by optical means. This opens up an entirely new avenue for the visualization of turbulence. Mapping the field of vorticity, not velocity, has advantages for both the classical and quantum ranges of the QT spectrum. Within the former (coarse-grained flow on lengthscales greater than the mean separation between vortex lines), the regions of enhanced vorticity, i.\,e. those with an enhanced density of vortex lines, will be most visible. Within the latter (small lengthscales that resolve discrete vortex lines), one will be able to observe such processes as vortex reconnections, Kelvin waves, and the emission and absorption of small vortex loops -- which are believed to be responsible for the quantum cascade of energy and control the dissipation of the vortex tangle \cite{KSPRB2008,KSPRL2008}. 

Micron-sized particles of solid hydrogen have already been used to tag vortex cores at high temperatures, $T \sim 2$\,K \cite{Lathrop,Sergeev09}; however, the invasive means of introduction and relatively large particle size  preclude implementation of this technique for low temperatures and small lengthscales.  
Potentially ideal tracers would be metastable molecules He$_2^*$ in the spin triplet state which have a relatively long lifetime of (13$\pm$2)\,s \cite{McKinsey99}. They can be created {\it in situ}  either by ionization in a strong laser field \cite{ApkarianLaser} or after recombination of injected negative and positive ions \cite{Tokaryk93}. Each molecule can be visualized many times by laser fluorescence \cite{McKinsey08}; and single-molecule resolution is, in principle, possible.  It was shown that the excimers can serve as tracers of the normal component in superfluid $^4$He at temperatures above 1\,K \cite{Mehrotra79}, and visualization has been successfully demonstrated \cite{Guo10}.  

In this Letter we show that He$^*_2$ molecules can be used as vorticity tracers in the $T = 0$ limit where there is no normal fluid. We demonstrate that they can be trapped on the vortex cores, measure the trapping diameter, and discuss the dynamics of the decorated QT. The possibility of trapping arises due to the strong repulsion from the outer electron of He$_2^*$:  the helium forms a bubble of radius $R_* \approx 7.0$\,\AA\ around the molecule \cite{Eloranta02,Guo}, which is attracted to the vortex core through the  Bernoulli pressure. The molecule's radius $R_*$ is comparable to that of a positive cluster-ion \cite{PositiveIons}, $R_+ \approx$ 7\,\AA, which  has the binding energy to a vortex of $\sim 20$\,K \cite{RD1969} and is known to be nearly permanently trapped by vortex lines at $T<0.6$\,K \cite{PositiveIons}. This suggests that He$_2^*$ molecules should also stay trapped at comparable temperatures. For a stationary bubble of volume $V=\frac{4\pi}{3}R_*^{3}$ at distance $r \gg R_*$ from a straight vortex, the interaction energy is \cite{RD1969, PZ1969, Thomson1873}
\begin{equation} 
U = - \frac{3\rho V \kappa^2}{16\pi^2}r^{-2}.
\label{Ur}
\end{equation} 
Numerical calculations by Schwarz \cite{Schwarz1974} showed that this formula works well for $\frac{r}{R} \geq 3$ and also for a bubble moving  with speeds of up to at least 5\,m\,s$^{-1}$. 
The singular potential $U = -k r^{-n}$ with $n=2$ is a special case between the less steep case $n<2$, that cannot capture particles of positive total energy in the absence of dissipation, and $n > 2$, where there is always a finite impact parameter within which the particle will be brought to the close vicinity of the singularity at $r=0$. For our case of $n=2$ a particle, arriving from infinity at speed $v$ with impact parameter $b$, is captured if $b \leq b_0(v)$ \cite{RD1969}, where  
\begin{equation}
b_0(v) = \left(\frac{2k}{Mv^2}\right)^{1/2} = \frac{\kappa}{2\pi} \left({\frac{3\rho V}{2M}}\right)^{1/2} v^{-1},
\label{b0}
\end{equation}
 and $M = \frac{\rho V}{2} + 2m_4 = 71{\rm \,amu}$ is the effective mass of the molecule $^4$He$_2^*$ (the product $\rho V$ is expected to be nearly pressure-independent because of  the opposite dependeces $\rho(P)$ and $V(P)$: between pressures $P=0.1$\,bar and $P=5.0$\,bar used in our experiment, the helium density $\rho$ increases by 5\% \cite{ElwellMeyer1967} while the bubble's volume $V \propto R_*^3$ is expected to decrease by 5\% \cite{Guo}). 
 In reality, the singularity in $U(r)$ at $r=0$ is replaced by a finite minimum of width $\sim R_*$, which would allow the particle eventually  to escape (effectively undergoing scattering by a large angle) -- provided little energy was dissipated on its way to $r \sim R_*$. However, the vortex lines in QT are neither rigid nor straight. The strong interaction with the vortex line at small $r$ might thus cause the dissipation of sufficient energy through the creation of Kelvin waves (as well as, perhaps, of phonons) -- this would result in a permanent trapping of the particle. Currently,  there is no theory of such processes applicable to excimers.  

\begin{figure}

\includegraphics[width=7cm]{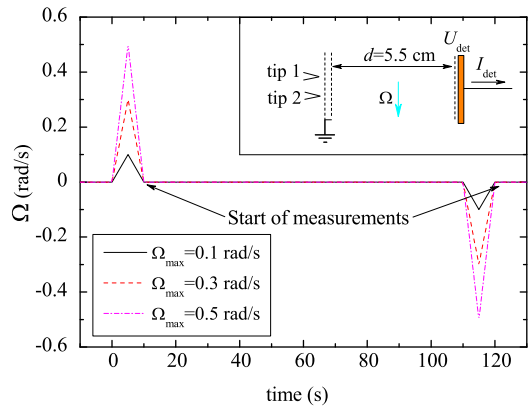}

\caption{(Color online) Time dependence of the cryostat angular velocity during the trapping diameter measurements for various agitation amplitudes $\Omega_\mathrm{max}$. Inset: sketch of the experimental setup and the direction of rotation.}
\label{omega_cell}
\end{figure}

 In  our experiments the molecules were produced through the application of a high voltage pulse (typically, -700\,V) to one of two sharp tungsten tips \cite{Golov98} (inset in Fig.~\ref{omega_cell}). Two grounded grids in front of the tips served as collectors of the electric charge. The corresponding tip currents were $\sim$ -1\,nA, and the pulses were 0.4\,s -- 2\,s long, causing heating by no more than a millikelvin. All experiments were performed at temperatures $T < 120$\,mK, where phonons are few and the molecules in the absence of vortices move ballistically with a broad spread of velocities centred at $\sim2$\,m\,s$^{-1}$ \cite{Zmeev12}. The detector  is a grid-shielded copper plate under an electric field strength of $\sim10^5$\,V\,cm$^{-1}$ situated $d=5.5$\,cm away from the tips. The molecules are ionized upon hitting the metal surface, resulting in a detectable electric current \cite{Surko68}. We have observed a slightly sublinear dependence of the detected current as a function of  the detector field and only a small difference for different signs of the electric field. In the experiments described the detector grid voltage was  $U_\mathrm{det}=-1200$\,V, so that the electrons in the main drift volume were forced away from the detector. Correspondingly, the electrons formed after ionization of the excimers were pushed towards the detector plate and the detected current  $I_\mathrm{det}$ had a negative sign. The efficiency of the detector is unknown, but smaller than unity, as we did not observe  saturation of the detected current on increasing the electric field. In these experiments it was essential to use isotopically ultrapure $^4$He \cite{Zmeev12, Hendry87}.


When ballistic molecules travel through a volume occupied by quantized vortices with vortex line density  $L$, some of them can get either trapped on the vortices or scattered from them. In this case the observed electric current $I$, due to a beam of excimers, will be attenuated:
\begin{equation}
I(L)=I(0)\mathrm{e}^{- \sigma  L d} ,
\label{attenuation}
\end{equation}            
where $\sigma$ is the trapping diameter, which is a characterization of the probability that the excimer does not arrive at the detector due to interaction with vortices. Regular arrays of rectilinear vortices with a known equilibrium line density $L_0=2 \Omega \kappa^{-1}$ could be created in the experimental volume by rotating the cryostat at a constant angular velocity $\Omega$.  In practice, however, the excimer signals were very weak and were degraded by noise introduced by the rotation. So instead of using uniform rotation, we created vortex tangles  inside the experimental cell by periodic rotational agitation of the cryostat \cite{WalmsleyAgitation} with amplitude $\Omega_\mathrm{max}$ and made measurements when the cryostat was stationary before the vortices had time to decay, as shown in Fig.~\ref{omega_cell}. The density of the vortices depends on $\Omega_\mathrm{max}$ and can readily be measured using charged vortex rings (CVRs) produced by the same tip with an electric field configuration similar to that used in \cite{Walmsley07,Walmsley08} for measurements of $L$. The trapping diameter for CVRs could be measured {\it in situ} using uniform rotation, as the CVR signals were several times  stronger than the excimer signals. It was found to be $\sigma_\mathrm{CVR}=(260\pm15)$\,nm for the experimental conditions (to facilitate CVR collection, the voltage between the injector grids and the detector was -30\,V). At $\Omega_\mathrm{max}=0.5$\,rad\,s$^{-1}$ we measured  $L$  to be 7.6$\times 10^3$\,cm$^{-2}$. 

\begin{figure}
\includegraphics[width=7cm]{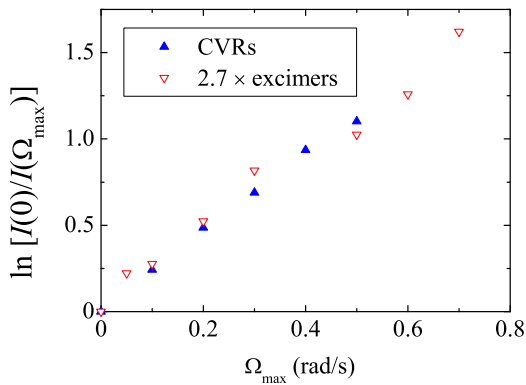}
\caption{(Color online) Attenuation of the signal from CVRs (solid triangles) and from excimers (open triangles) due to the presence of quantized vortices in the experimental volume after the agitation. The data for the excimers is multiplied by a factor of 2.7.  Each point presented is the result of averaging over $\sim$100 pulses.  $P=0.1$\,bar, $T=60$\,mK. }
\label{lnI}
\end{figure}

By comparing the attenuation of the excimer signal, and the CVR signal, for CVRs with a known trapping diameter after  identical agitation of the superfluid, we can calculate $\sigma$ using formula~(\ref{attenuation}). Fig.~\ref{lnI} shows that the attenuation of the excimers was on average 2.7 times weaker; hence the trapping diameter was
$\sigma=(96\pm 6)$\,nm at a pressure of $P=0.1$\,bar and $T=60$\,mK.  A similar value of $\sigma \approx 100$\,nm was also measured at 100\,mK. At 50\,mK we attempted a measurement in uniform rotation at $\Omega=2.5$\,rad\,s$^{-1}$ and within the error obtained the same value.  However, at a higher pressure of 5.0\,bar and at $T=60$\,mK we found $\sigma=(27\pm5)$\,nm.  


 For the mean experimental velocities $v=1.7 (2.4)$\,m\,s$^{-1}$, at $P=0.1(5.0)$\,bar \cite{Zmeev12}, Eq.\,\ref{b0} yields the capture diameter of $2b_0 = 31 (22)$\,nm. 
  For the impact parameter $b>b_0$, the scattering angle $\chi (v,b)$ is 
\begin{equation}
\chi =  \pi + \int_{\infty}^{r_{\rm m}}\frac{{\rm d}r}{r(r^2-r_{\rm m}^2)^{1/2}} = \pi\left[\left(1-\frac{b_0^2}{b^2} \right)^{-1/2}-1\right]
\label{chi}
\end{equation}
(here $r_{\rm m} = \sqrt{b^2-b_0^2}$ is the distance of closest approach of molecule to the vortex core). If the molecules are injected in a large angle, for which we have a strong evidence, the effective reduction of the signal would occur at $|\chi| \sim \pi/2$, as scattering by small angles will not change the total signal. The result $ 2b (\pi/2)  = 2\times \frac{3}{\sqrt{5}}b_0 = 41 (29)$\,nm
is a factor of two smaller than the experimental $\sigma = 96$\,nm at $P=0.1$\,bar, but  close to $\sigma = 27$\,nm at $P=5$\,bar. This formula gives the lower estimate for $\sigma$ as any dissipative mechanisms (such as generation of Kelvin waves on vortices passed by the molecule) should increase the  value for $\sigma$. Furthermore, in the estimates we used mean velocities for the excimers, while in reality the widths of the velocity distributions were not small~\cite{Zmeev12}.



So far we have assumed that the excimers travel as free molecules. They might in principle be trapped on vortex rings. These  rings could either be nucleated during the  process of molecule creation or be emitted from the dense  tangle near the tip. 
 The self-induced velocity $v_{\rm v}$ 
of a vortex ring of radius  $R_{\rm v}$ is
 $$v_{\rm v} \approx \frac{\kappa}{4\pi R_{\rm v}}\ln\frac{8R_{\rm v}}{a_0},$$ 
where $a_0 = 0.81$\,\AA($0.86$\,\AA) \cite{GlabersonDonnelly1986, RR1964,Steingart1972}. 
  The mean experimental velocity $v=1.7(2.4)$\,m\,s$^{-1}$ corresponds to the radii $R_{\rm v} = 36(24)$\,nm 
for the alleged rings. For CVRs interacting with smooth vortex lines, the effective interaction diameter is nearly geometrical, $\sigma_{\rm v} \approx 2 R_{\rm v}$ \cite{SchwarzDonnelly1966}. Applying the same relation to the alleged ring radii $R_{\rm v}$, we thus arrive at the expected trapping diameters of $\sigma_{\rm v} = 72$\,nm (48\,nm) at $P=0.1$\,bar (5.0\,bar). When compared with experimental $\sigma = 96$\,nm (27\,nm), these are not too far off. Yet, we would treat the model of ring-bound molecules as less likely. 
It is inconsistent with our measurements of the time of flight at higher temperatures reported in \cite{Zmeev12}: as the drag due to  interactions with phonons increases with temperature, the rings are expected to shrink and move faster; however, the experimental values of the mean velocity were almost temperature-independent.

\begin{table}[h]
\begin{center}
\begin{tabular}{|c|c|c|}
\hline  
$P$ & 0.1\,bar & 5.0\,bar \\ \hline 
$\sigma$ & $96\pm 6$\,nm & $27\pm 5$\,nm \\ 
$2 b_0$ & 31\,nm & 22\,nm \\
$2 b \left( \frac{\pi}{2} \right)$ & 41\,nm & 29\,nm \\ 
$2 R_{\rm v}$ & 72\,nm & 48\,nm \\ \hline
\end{tabular}
\end{center}
\caption{Experimental values of the trapping diameter, $\sigma$, along with theoretical estimates for capture, $2 b_0$, and scattering, $2 b$, by $\chi = \pi/2$ of a bare molecule, and scattering of a molecule riding on a vortex ring $2 R_{\rm v}$.}
\label{tab}
\end{table}

  

\begin{figure}

\includegraphics[width=7cm]{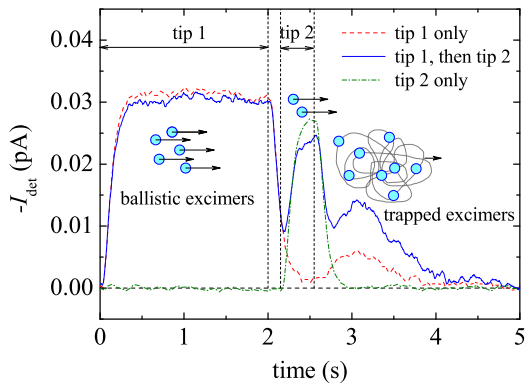}

\caption{(Color online) Experiments demonstrating decoration of vortices. Dashed red signal: After a 2\,s current pulse from tip 1 a slow-moving tangle of vortices was formed. Solid blue signal: a short current pulse from tip 2 followed the  same pulse from  tip 1, the slow peak increased in amplitude, but neither its time of flight nor its width had changed. Dash-dotted green signal: when the short pulse from tip 2 was not preceded by the pulse from tip 1, the fast signal was larger in amplitude. $P=0.1$\,bar, $T=60$\,mK. Each curve is a result of averaging over $\sim$100 signals.}
\label{Decoration}
\end{figure}

 Our experimental values of $\sigma$ are compared with the theoretical estimates in Table\,\ref{tab}. Further experiments showed that molecules are emitted in a large solid angle, so we believe that the experimental $\sigma$ mainly quantifies the trapping processes and that the enhancement by the large-angle scattering is not substantial. To demonstrate that the excimers can be captured and transported by the vortices, we performed an experiment where the vortices were produced by ion injection rather than rotation. In the experiments described so far mainly ballistic excimers were produced by applying a short (500\,ms) pulse to one tip. However,  intensive and long enough pulses can produce a vortex tangle slowly moving from the tip towards the detector \cite{Walmsley10}. As is shown by the broken red line in  Fig.\,\ref{Decoration}, most of the excimers produced in a 2\,s long pulse still arrive at the detector promptly,  corresponding to ballistic  propagation, but a relatively small fraction arrive in a broad pulse with a delay of roughly 1\,s. (The amplitude of the delayed pulse scaled with the electric field in the detector as it also does for the ballistic signal. This suggests that it was still molecules, rather than some other particle or excitation, that  were being detected.)  We interpret this delayed pulse as being due to molecules that have been trapped on the vortex tangle,  which drifts slowly towards the detector, where the arrival of the molecules is registered (tangles of similar densities do not decay much in times of order 1\,s \cite{Walmsley08}).

  In principle, the slow pulse might have been due to the formation of a group of very slowly moving,  but untrapped,  molecules. That this is not the case is shown by the results of a second experiment in which tip 2 is used to create,  with a short pulse, a second group of molecules.  The dash-dotted green line shows that pulsing  tip 2 alone in this way produces no delayed pulse;  this is because a short pulse does not produce a drifting vortex tangle.  But when  tip 2 is pulsed shortly after tip 1 was pulsed (the precise timing is unimportant),  we get the solid blue line,  which shows that (a) fewer of these extra molecules reach the detector very quickly, and (b) there is an enhancement of the delayed signal due to the first pulse by a factor of about 2.5,  with no alteration in its time of flight or its width.  Thus some of the molecules produced by tip 2 must have simply joined the collection of molecules already trapped on the vortex tangle,  without changing its characteristics. It is significant  that the enhancement is greater than the decrease in the amplitude of the fast signal (from tip 2);  this can be the case only if molecules were emitted from the tips into a wide solid angle, allowing the tangle to pick up more molecules than would have originally been heading for the detector.  The fact that  there can be such a large enhancement of the slow signal  suggests that molecules,  once trapped,  do not easily escape from the tangle,  and do not interact in such a way that there is rapid annihilation through Penning ionization. Using the measured $\sigma$ and the attenuation of the second pulse,  we can estimate that the line density in the slow-moving tangle is $ L\sim4\times 10^3$\,cm$^{-2} $ and,  with the assumption that the detector is 100\,\% efficient, that the mean separation between trapped molecules is about 1\,mm.  However, the efficiency of a similar detector used in \cite{Surko68} was $\lesssim 1$\,\%, which lowers the mean separation of trapped excimers to $\lesssim 10$\,$\mu$m.

To conclude, we have shown that the \exc~molecules at low temperatures are trapped on vortices and hence can be used as tracers for visualization of  quantum turbulence in the zero temperature limit in an optical experiment utilizing induced fluorescence. The measured trapping diameter is $\sigma \sim 100$\,nm; such a large value is most likely related to the effectiveness of the $U\propto -r^{-2}$ potential in capturing slow particles at relatively large impact parameters. The large value of the trapping diameter, in turn, guarantees that the concentrations of excimers, required to decorate vortex lines, can be produced in superfluid helium {\it in situ} without significant overheating. The observed $\sim 3.5$-fold reduction in $\sigma$ between pressures 0.1\,bar and 5\,bar, stronger than that expected from our theory, requires further investigation of this phenomenon in a broader range of pressures.


This work was supported through the Materials World Network program by the Engineering and Physical Sciences Research Council [Grant No. EP/H04762X/1] and the National Science Foundation [Grants DMR-1007937 and DMR-1007974]. PMW is indebted to EPSRC for the Career Acceleration Fellowship [Grant No. EP/I003738/1].



\begin{thebibliography}{99}

\bibitem{Vinen06} W. F. Vinen, J. Low Temp. Phys. 145, 7 (2006).

\bibitem{Nobach2009} H. Nobach, E. Bodenschatz , Experiments in Fluids {\bf 47}, 27 (2009). 

\bibitem{KSPRB2008} E. V. Kozik and B. V. Svistunov, Phys. Rev. B {\bf 77}, 060502(R) (2008).

\bibitem{KSPRL2008} E. V. Kozik and B. V. Svistunov, Phys. Rev. Lett. {\bf 100}, 195302 (2008).

\bibitem{Lathrop} G. P. Bewley, D. P. Lathrop, and K. R. Sreenivasan, Nature {\bf 441}, 588 (2006).

\bibitem{Sergeev09} Y. A. Sergeev and C. F. Barenghi, J. Low Temp. Phys. {\bf 157}, 429 (2009).

\bibitem{McKinsey99} D. N. McKinsey, C. R. Brome, J. S. Butterworth, S. N. Dzhosyuk, P. R. Huffman, C. E. H. Mattoni, J. M. Doyle, R. Golub and K. Habicht, Phys. Rev. A {\bf  59}, 200 (1999).


\bibitem{ApkarianLaser} A. V. Benderskii, R. Zadoyan, N. Schwentner, and V. A. Apkarian, J. Chem. Phys. {\bf 110}, 1542 (1999).

\bibitem{Tokaryk93} J. W. Keto, M. Stockton, and W. A. Fitzsimmons, Phys. Rev. Lett. {\bf 28}, 792 (1972); J. W. Keto, F. J. Soley, M. Stockton, and W. A. Fitzsimmons, Phys. Rev. A {\bf 10}, 872 (1974); D. W. Tokaryk, R. L. Brooks, and J. L. Hunt,  Phys. Rev. A {\bf 48}, 364 (1993).

\bibitem{McKinsey08}W. G. Rellergert, S. B. Cahn, A. Garvan, J. C. Hanson, W. H. Lippincott, J. A. Nikkel, and D. N. McKinsey, Phys. Rev. Lett. {\bf 100}, 025301 (2008);  D. N. McKinsey, W. H. Lippincott, J. A. Nikkel, and W. G. Rellergert, Phys. Rev. Lett. {\bf 95}, 111101 (2005).

\bibitem{Mehrotra79} R. Mehrotra, E. K. Mann and A. J. Dahm, J. Low Temp. Phys. {\bf 36}, 47 (1979).

\bibitem{Guo10} W. Guo, S. B. Cahn, J. A. Nikkel, W. F. Vinen, D. N. McKinsey, Phys. Rev. Lett. 105, 045301 (2010).

\bibitem{Eloranta02} J. Eloranta, N. Schwentner, V. A. Apkarian, J. Chem. Phys. {\bf 116}, 4039 (2002).

\bibitem{Guo} The excimer bubble's radius $R_*$  only weakly depends on pressure because the energy and radius of the outer electron is mainly controlled by the strong Coulomb attraction to the centre of the molecule. The calculations by Wei Guo (unpublished), in the framework of continuum model, reveal that between 0 and 5\,bar $R_*^3$ decreases by less than 5\%. 

\bibitem{PositiveIons} G. A. Williams, K. DeConde, and R. E. Packard, Phys. Rev. Lett. {\bf 34}, 924 (1975).

\bibitem{RD1969} R. J. Donnelly and P. H. Roberts, Proc. Roy. Soc. London, Series A, {\bf 312}, 519 (1969).

\bibitem{PZ1969} W. P. Pratt, Jr., W. Zimmermann, Jr., Phys. Rev. 177, 412 (1969)

\bibitem{Thomson1873} The original formula used by Roberts and Donnelly is a factor of 1/3 smaller as it does not take into account the distrotion of the flow around the sphere (calculated by W. Thomson, Phil. Mag. {\bf 45}, 332 (1873)). 

\bibitem{Schwarz1974} K. W. Schwarz, Phys. Rev. A {\bf 10}, 2306 (1974).

\bibitem{ElwellMeyer1967} D. L. Elwell and H. Mayer, Phys. Rev. {\bf 164}, 245 (1967).

\bibitem{Golov98} A. Golov and H. Ishimoto, J. Low Temp. Phys. {\bf 113}, 957 (1998).

\bibitem{Zmeev12} D. E. Zmeev, F. Pakpour, P. M. Walmsley, A. I. Golov, P. V. E. McClintock, S. N. Fisher, W. Guo, D. N. McKinsey, G. G. Ihas and W. F. Vinen,
J. Low Temp. Phys.  {\bf 171},  207 (2013).

\bibitem{Surko68} C. M. Surko and F. Reif,  Phys. Rev. {\bf 175}, 229 (1968).

\bibitem{Hendry87} P. C. Hendry and P. V. E. McClintock, Cryogenics 27, 131 (1987).


\bibitem{WalmsleyAgitation} P. M. Walmsley and A. I. Golov, Phys. Rev. B {\bf 86}, 060518(R) (2012).

\bibitem{Walmsley07} P. M. Walmsley, A. I. Golov, H. E. Hall, A. A. Levchenko, and W. F. Vinen, Phys. Rev. Lett. {\bf 99}, 265302 (2007).

\bibitem{Walmsley08} P. M. Walmsley and A. I. Golov, Phys. Rev. Lett. {\bf 100}, 245301 (2008).

\bibitem{GlabersonDonnelly1986} W. I. Glaberson and R. J. Donnelly, Structure, distribution and dynamics of vortices in helium II in Progress in Low Temperature Physics IX (D. F. Brewer, ed.) North-Holland, Amsterdam (1986). 

\bibitem{RR1964} G. W. Rayfield and F. Reif, Phys. Rev. {\bf 136}, A1194 (1964).

\bibitem{Steingart1972} M. Steingart, W. I. Glaberson, J. Low Temp. Phys. {\bf 8}, 61 (1972). 

\bibitem{SchwarzDonnelly1966} K. W. Schwarz and R. J. Donnelly, Phys. Rev. Lett. {\bf 17}, 1088 (1966).

\bibitem{Walmsley10} A. I. Golov, P. M. Walmsley, P. A. Tompsett, J. Low Temp. Phys. {\bf 161}, 509 (2010).



\end{thebibliography}
\end{document}